\documentclass[aps,a4paper,twoside,twocolumn,secnumarabic,balancelastpage,amsmath, amssymb,superscriptaddress]{revtex4-2}
\usepackage[pdftex]{graphicx}
\usepackage{longtable}
\usepackage{bm}
\usepackage{bbold}
\usepackage{verbatim}
\usepackage{mathtools}
\usepackage{adjustbox}
\usepackage[colorlinks=true]{hyperref}
\usepackage{xcolor}
\usepackage{todonotes}

\begin{document}

\title{Electronic and structural properties of crystalline and amorphous (TaNbHfTiZr)C from first principles}
\date{\today}

\author{Bram van der Linden}
\affiliation{Institute of Physics, University of Amsterdam, 1098 XH Amsterdam, The Netherlands}
\author{Tadeus Hogenelst}
\affiliation{Advanced Research Center for Nanolithography, ARCNL, Science Park 106, Amsterdam, Netherlands}
\author{Roland Bliem}
\affiliation{Institute of Physics, University of Amsterdam, 1098 XH Amsterdam, The Netherlands}
\affiliation{Advanced Research Center for Nanolithography, ARCNL, Science Park 106, Amsterdam, Netherlands}
\author{Kateřina Dohnalová}
\affiliation{Institute of Physics, University of Amsterdam, 1098 XH Amsterdam, The Netherlands}
\author{Corentin Morice}
\affiliation{Institute for Theoretical Physics and Delta Institute for Theoretical Physics, University of Amsterdam, 1090 GL Amsterdam, The Netherlands}
\affiliation{Laboratoire de Physique des Solides, CNRS UMR 8502, Université Paris-Saclay, F-91405
Orsay Cedex, France}

\begin{abstract}
High entropy materials (HEMs) are of great interest for their mechanical, chemical and electronic properties. In this paper we analyse (TaNbHfTiZr)C, a carbide type of HEM, both in crystalline and amorphous phases, using density functional theory (DFT). We find that the relaxed lattice volume of the amorphous phase is larger, while its bulk modulus is lower, than that of its crystalline counterpart. Both phases are metallic with all the transition metals contributing similarly to the density of states (DOS) close to the Fermi level, with Ti and Nb giving the proportionally largest contribution of states. We confirm that despite its great structural complexity, 2x2x2 supercells are large enough for reliable simulation of the presented mechanical and electronic properties by DFT.      

\end{abstract}

\maketitle
\section{Introduction}
Disorder in solids can be present in various forms. Usually, the concept is associated with structural disorder, which starts with defects in single crystals and culminates in liquid-like, glassy assemblies of atoms. The lack of long-range structural order in amorphous solids causes the electronic structure of a material, as well as the mechanical, optical, thermal or magnetic properties to differ notably from those of crystalline solids \cite{suryanarayana, trexler_bulkmetallic_2010, hunderi_amorphousmetalfilms_1976, stachurski_structpropamorphous_2011}. In particular, amorphous alloys show superior corrosion resistance \cite{scully_corrosion_2007, asami_corrosionresistance_1976}, mechanical strength \cite{chen_glassymetals_1980, gilman_metallicglasses_1980}, and low permeability for diffusion \cite{barbour_selfdiffusion_1985}. 

Another type of disorder that can strongly affect the properties of a solid is configurational disorder, which is best illustrated by the example of the novel class of high-entropy materials (HEMs), compounds with five or more principal elements \cite{cantor_multicomponentalloys_2004,yeh_nanostructured_2004}.
Distributing this large number of constituent elements among well-defined positions of a crystalline lattice allows the material to maximize its configurational entropy, resulting in a tendency to form a single-phase solid solution. 
The properties of HEMs have been shown experimentally to be superior to the superposition of its constituents \cite{braic_nanostructured_2012, castle_processing_2018}. The remarkable effect of configurational disorder was first demonstrated for metallic alloys \cite{cantor_multicomponentalloys_2004, yeh_nanostructured_2004}, but further extends to ceramic materials (e.g. carbides, oxides or silicides) \cite{oses_high-entropy_2020}.
High-entropy ceramics typically consist of two sublattices: an ionic or covalent backbone (e.g. C, O, or Si) with an interstitial lattice occupied by five or more principal elements, typically metals.
For example, in a rocksalt high-entropy metal carbide (HEMC) all metals occupy the central sites in octahedra of an fcc lattice of carbon, but their arrangement on this sublattice is random. 
Experimental and computational studies \cite{oses_high-entropy_2020} suggest that HEMCs have excellent application-oriented properties. One such example is  (TaNbHfTiZr)C that exhibits a great combination of phase purity and stability \cite{george_highentropyalloy_2019, harrington_phasestability}, as well as predictions of enhanced hardness \cite{sarker_high-entropy_2018, wang_enhancedhardness_2020} and low thermal conductivity \cite{yan_hf_2018}.

In addition to the effects of configurational disorder, structural disorder is recognized to have a strong influence on materials properties and is often used intentionally to adjust the performance of a material to its application. In cases of materials without configurational disorder, such as, e.g., structurally disordered transition metal carbides,  tuneable \cite{sanchez_lopez_tribological} tribological properties have been demonstrated, as well as functionality for energy storage applications \cite{bi_ZrC-aC_2016}. 
In many cases such disordered carbides are not amorphous at the atomic scale but have a composite structure of nanocrystalline metal carbide particles embedded in amorphous graphite \cite{jansson_TMCfilms_2013, folkenant_amorphouscarbide_2015, nedfors_nanocrystallineNbC_2011, zhao_corrosionResistanceCrC_2018, magnuson_TiC_2009}. 
The addition of weak carbide-forming alloying elements aids in the full amorphization of these materials \cite{jansson_TMCfilms_2013}. Additionally, electronic and mechanical properties of amorphous transition metal carbides have shown to depend significantly on the bond distribution between C and metals \cite{jansson_TMCfilms_2013, jansson_Nb-Si-C_2013} and elastic moduli of amorphous coatings may be modified by changing the transition metal M \cite{jansson_Nb-Si-C_2013}. 
Compared to a single-crystalline structure, amorphous transition metal carbide coatings are considered less prone to cleavage fracture and decohesion of grain boundaries, reducing the brittleness of a functional coating by reduction of average bond energy \cite{kaloyeros_amorphous_1986}. 
This indicates that modifying the level of disorder is a pathway to tune electronic and mechanical properties in transition metal carbide coatings. Disorder is thus highly relevant in applications, and the relation between disorder and material properties calls for its introduction as a design parameter. 

In this work we investigate the effects of disorder on the mechanical properties and the electronic structure of high-entropy refractory metal carbides. Both configurational and structural disorder are introduced by studying the HEMC (TaNbHfTiZr)C in its crystalline and fully amorphous form at identical compositions. We approach the challenge to model amorphous systems from first principles using stochastic quenching \cite{kadas_structural_2012} and density functional theory (DFT) \cite{yang_structural_2018}.\newline

\begin{figure*}
\centering
\begin{tabular}{crcrl} 
(a) & \raisebox{-0.95\height}{\includegraphics[width=0.25\textwidth]{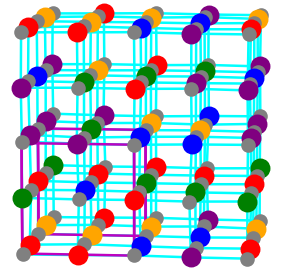}} &
(b) & \raisebox{-0.95\height}{\includegraphics[width=0.25\textwidth]{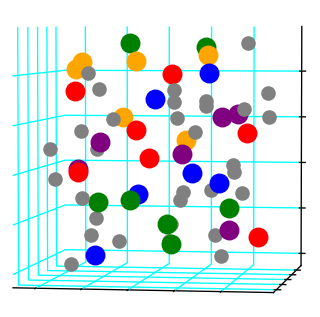}} & 
\raisebox{-1.1\height}{\includegraphics[width=0.10\textwidth]{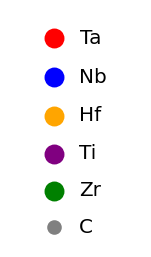}}
\end{tabular}
\caption{Crystal structures of a crystalline supercell (a) and its amorphous counterpart (b) with the same stoichiometry. Both are relaxed with DFT. The magenta cube in (a) represents the rock-salt unit cell, the distance between two grid lines in (b) is 2 \AA.}
\label{crystalline_amorphous_visualization}
\end{figure*}

\section{Methods}
For convenience, in this paper we use the same naming convention for (TaNbHfTiZr)C as in \cite{yang_structural_2018}.
The crystal structure of crystalline (TaNbHfTiZr)C was modelled using $2\times2\times2$ supercells in the rock-salt structure (Fig.~\ref{crystalline_amorphous_visualization} (a)), following \cite{yang_structural_2018}. Each supercell includes 32 transition metal atoms, and can therefore not be fully stoichiometric. For this reason we analyse several samples with small variations to the stoichiometry, to study how small variations in composition influence material properties. For this we randomly generated nine supercells with slightly different stoichiometries (see Table \ref{stoichiometries}), and also ran calculation on the supercell used in \cite{yang_structural_2018}, for a total of 10 supercells.

To analyse the additional effects of structural disorder, we also generated ten amorphous supercells with exactly the same stoichiometry as their crystalline counterparts (e.g. Fig.~\ref{crystalline_amorphous_visualization}(b)). Amorphous structures were obtained using stochastic quenching \cite{holmstrom_structure_2010}, as follows: first, atoms were placed randomly in a cubic supercell with periodic boundary positions and a side length $L=9.04$ \AA, which is twice the lattice constant. To make sure that we keep the energy of supercells reasonably low, we constrained the distance between every two atoms to be larger than 0.4 \AA \ \cite{kadas_structural_2012}. Second, we relaxed the atomic coordinates. To lower the cost of the \emph{ab-initio} calculations, we used first a low-precision method ---a steepest descent algorithm and a Mie potential \cite{GonzalezLopez2021}--- to minimize the force on each atom, and set the convergence criterion to $10^{-10}\times f_{0}$ where $f_{0}$ is the typical force, defined as
\begin{align}
f_{0}=\sqrt{\frac{1}{N}\sum_{k} \left( \frac{\partial U_{LJ}(r_{k})}{\partial r_{k}}\right)^{2}}
\label{f0}
\end{align}
with $r_{k}$ the distance between two particles of pair $k$ and $N$ the number of atoms in the amorphous supercell. 

Final relaxation of the unit cell size and atomic coordinates, as well as calculation of the electronic properties, were performed using DFT, as implemented in CP2K \cite{kuhne_cp2k_2020} using the Perdew-Burke-Ernzerhof (PBE) exchange-correlation functional \cite{perdew_generalized_1996}, 
the Goedecker-Teter-Hutter (GTH) pseudopotentials \cite{goedecker_separable_1996} and short range Gaussian basis functions \cite{vandevondele_gaussian_2007}. Four real-space grids were used, with a cut-off energy of the finest grid of 900 Ry and a cut-off controlling the distribution of products of Gaussians on the different grids of 60 Ry. Reciprocal space was sampled using a $3\times3\times3$ Monkhorst-Pack grid \cite{monkhorst_special_1976}. All the parameters were thoroughly tested for convergence.

Atomic positions and lattice parameters were relaxed using the conjugate gradient algorithm. The convergence criteria for the maximum and the root-mean square atomic force were set to 0.001 Ha/Bohr, and the ones for the difference with respect to the previous optimization step were set to 0.001 Bohr. In every crystalline and amorphous supercell, the atomic positions and cell size were relaxed at the same time, and the cells were required to remain cubic. Additionally, the pressure was converged to be $(1\pm 10)$ atm after relaxation. Density of states (DOS) and bulk modulus calculations were performed at zero pressure.

Bulk moduli at zero-pressure were calculated by fitting the total energy as a function of the unit cell volume to the third-order Birch-Murnaghan equation of state \cite{birch_finite_1947, murnaghan}:
\begin{align}
E(V)={}&E_{0}+\frac{9V_{0}B_{0}}{16}\left\{    \left[\left(\frac{V_{0}}{V}\right)^{2/3}-1\right]^{3}B_{0}'\right. \nonumber
\\
& \left. +\left[\left(\frac{V_{0}}{V}\right)^{2/3}-1\right]^{2}   \left[6-4\left(\frac{V_{0}}{V}\right)^{2/3}\right]  \right\}
\label{BM}
\end{align}
where $E$ is the total energy of the supercell, $V$ is the supercell  volume, $E_{0}$ the equilibrium energy, $V_{0}$ the equilibrium volume, $B_{0}$ the zero-pressure bulk modulus, and $B_{0}'$ its pressure derivative. Each supercell was set to five or seven different volumes around the  equilibrium volume, to obtain a good fit to eq.~\ref{BM}.

We also calculated the DOS of the relaxed supercells. CP2K can only perform DOS calculations at the $\Gamma$-point, therefore we repeated the supercell three times in each Cartesian directions, which we found to be well converged. We used a Gaussian smearing with a standard deviation of $0.001$ Ha.

\section{Results}

\subsection{Structural properties}

The DFT-relaxed lattice constant for one rock-salt unit cell in the crystalline supercell was calculated to be $(4.519 \pm 0.006)$ \AA. This is very close to the experimentally measured values 4.5084 \AA \cite{zhou_high-entropy_2018}, 4.5180 \AA \cite{yan_hf_2018}, 4.497 \AA \cite{ye_firstprinciples_2019}, 4.500 \AA \cite{sarker_high-entropy_2018} and $(4.60\pm 0.3)$ \AA \cite{braic_nanostructured_2012}. The uncertainty found in this work is small compared to the spread in the reported experimental values, where the smallest and largest lattice constant differ by 0.103 \AA. In addition, the variation between supercells is of the order of the typical variation in relaxed lattice parameters between different DFT codes \cite{lejaeghere_reproducibility_2016}, pointing towards a low sensitivity of the relaxed lattice constant to the variations in the crystal's stoichiometry.

To be able to quantify the influence of the stoichiometry on the relaxed lattice constant, we consider the binary carbides TaC, NbC, HfC, TiC and ZrC which crystallize into rock-salt structures and can be considered as ``building blocks'' for the (TaNbHfTiZr)C crystal. For each of these binary carbides, the average volume per transition metal and carbon atom is $\left(a_{TMC}\right)^{3}/4$ where $a_{TMC}$ is the lattice constant of the binary carbide. So, for each supercell, the binary carbides (BC) can be used to define a volume 
\begin{align}
V_{BC}=\frac{1}{4}\sum_{TM}N_{TM}\cdot \left(a_{TMC}\right)^{3}
\label{volume_correlation_formula}
\end{align}
where $TM \in \left\{ \text{Ta}, \text{Nb}, \text{Hf}, \text{Ti}, \text{Zr} \right\}$ and $N_{TM}$ is the number of transition metals in a supercell. For each transition metal species, we calculated $V_{BC}$ using the lattice constants $a_{TMC}$ in \cite{yang_structural_2018}.

\begin{figure*}
\centering
\begin{tabular}{crcr}
(a) & \raisebox{-0.95\height}{\includegraphics[width=0.45\textwidth]{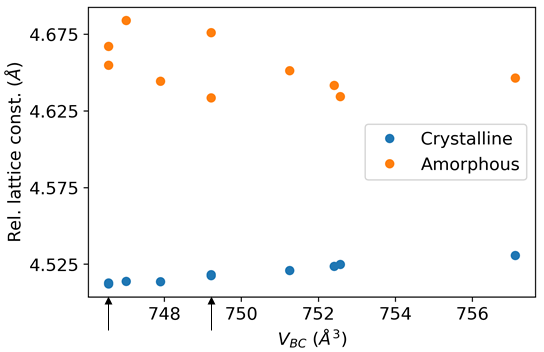}} &
(b) & \raisebox{-0.95\height}{\includegraphics[width=0.45\textwidth]{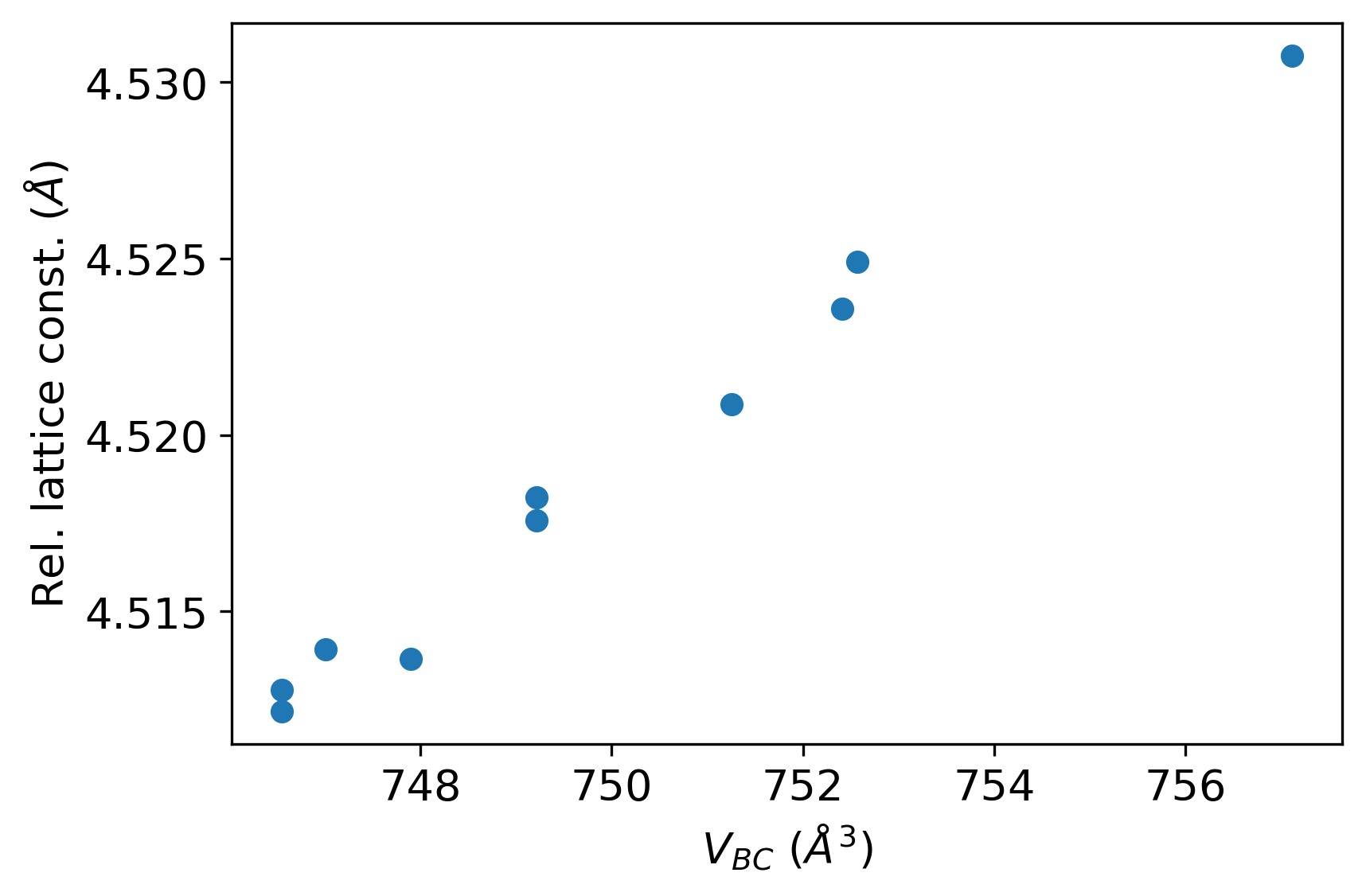}}
\end{tabular}
\caption{(a) Calculated relaxed lattice constant of the crystalline and amorphous supercells as a function of $V_{BC}$, as defined in Eq.~\eqref{volume_correlation_formula}, where the lattice constants of elemental carbides are taken from \cite{yang_structural_2018}. Arrows indicate supercells with the same stoichiometry. Note that only see one blue point is visible in these two cases, because the results are too close. (b) Repeated data from (a) for the crystalline case, showing a clear linear trend as a function of $V_{BC}$, indicating that the $V_{BC}$ is a good quantification parameter for the stoichiometry's role in lattice constant size.}
\label{lattice_constant_volume_without_config1_5_combined}
\end{figure*}

Fig.~\ref{lattice_constant_volume_without_config1_5_combined}a displays the relaxed lattice constants of the crystalline and amorphous supercells as a function of $V_{BC}$. The amorphous lattice constant was found to be $(4.65 \pm 0.02)$ \AA, about $0.15$ \AA \ larger than for the crystalline cells, exceeding the standard deviation due to the changes in configuration and stoichiometry. In addition, we observe a clear linear trend in the crystalline case, showing that the stoichiometry dependence is well quantified by $V_{BC}$. This trend is not clearly visible in the amorphous case, which could be related to the fact that the volume is larger because of the amorphous configuration, hence the different sizes of the metallic elements play a smaller part.

Out of the ten (crystalline or amorphous) supercells, as a control sample, we designed two pairs of supercells to have the same stoichiometry and hence the same $V_{BC}$. Fig.~\ref{lattice_constant_volume_without_config1_5_combined}b shows that the crystalline lattice constants corresponding to these supercells are remarkably close to each other, compared to the other supercells. However this is not the case in the amorphous case, as displayed in Fig.~\ref{lattice_constant_volume_without_config1_5_combined}a.

\subsection{Mechanical properties}

\begin{figure}
\centering
\includegraphics[width=0.45\textwidth]{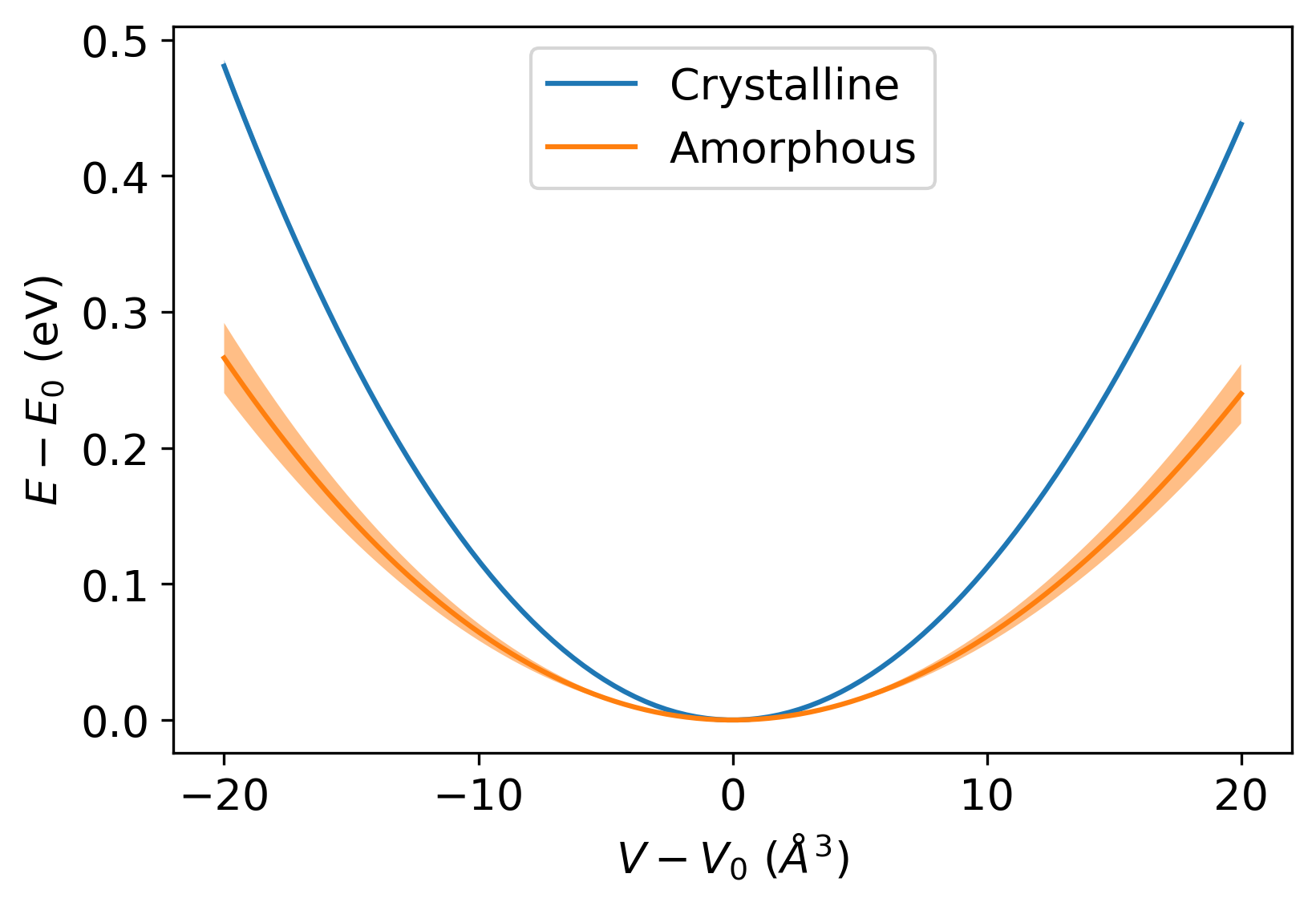}
\caption{Average fit of the total energy of relaxed supercells as a function of their volume to the third-order Birch-Murnaghan equation of state. The standard deviation corresponds to the width of the colored area around the curve. In the crystalline case, the standard deviation is too small to be distinguished.} 
\label{E_vs_V_all_supercells_combined}
\end{figure}

\begin{figure*}
\centering
\begin{tabular}{crcr}
(a) & \raisebox{-0.95\height}{\includegraphics[width=0.45\textwidth]{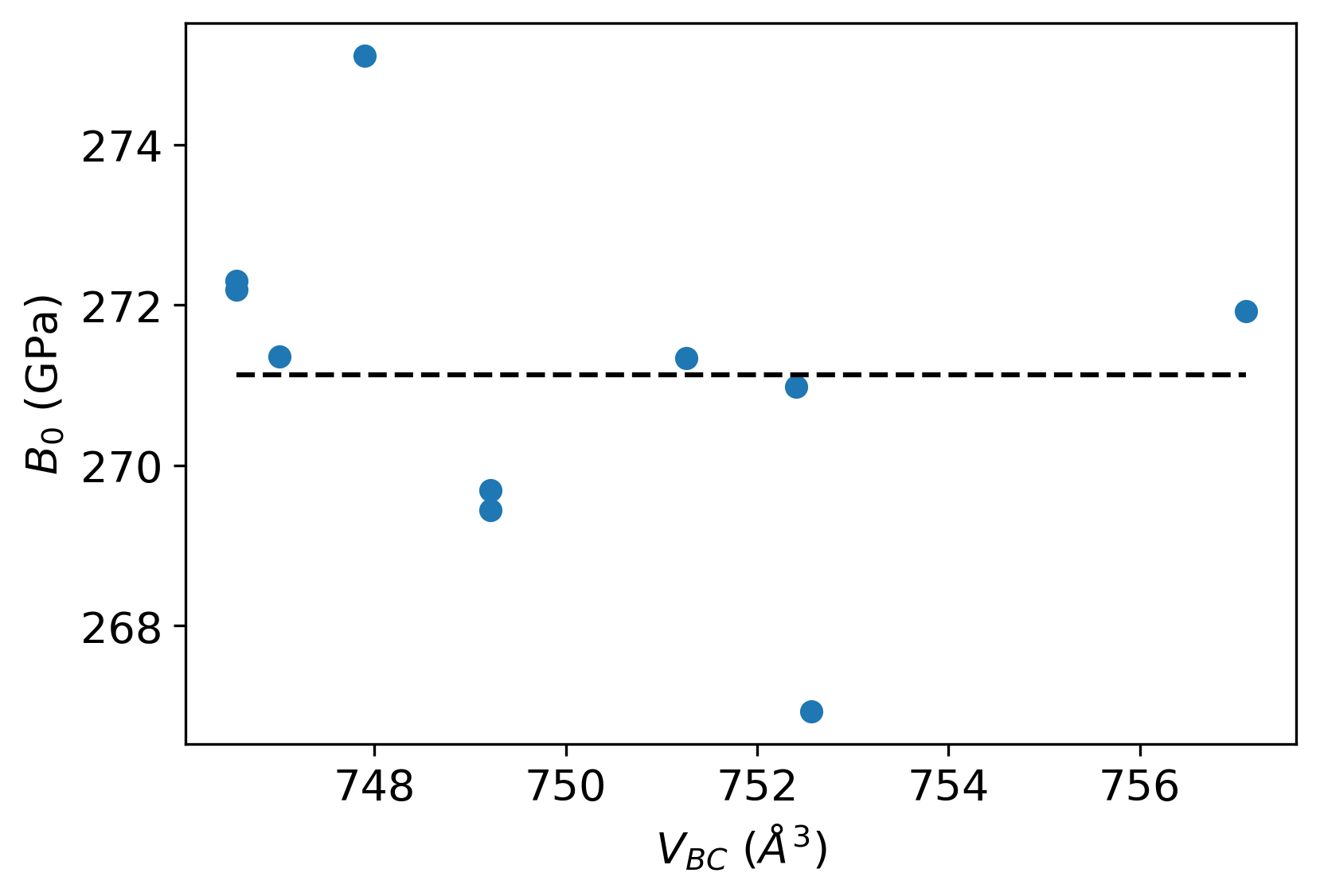}} &
(b) & \raisebox{-0.95\height}{\includegraphics[width=0.45\textwidth]{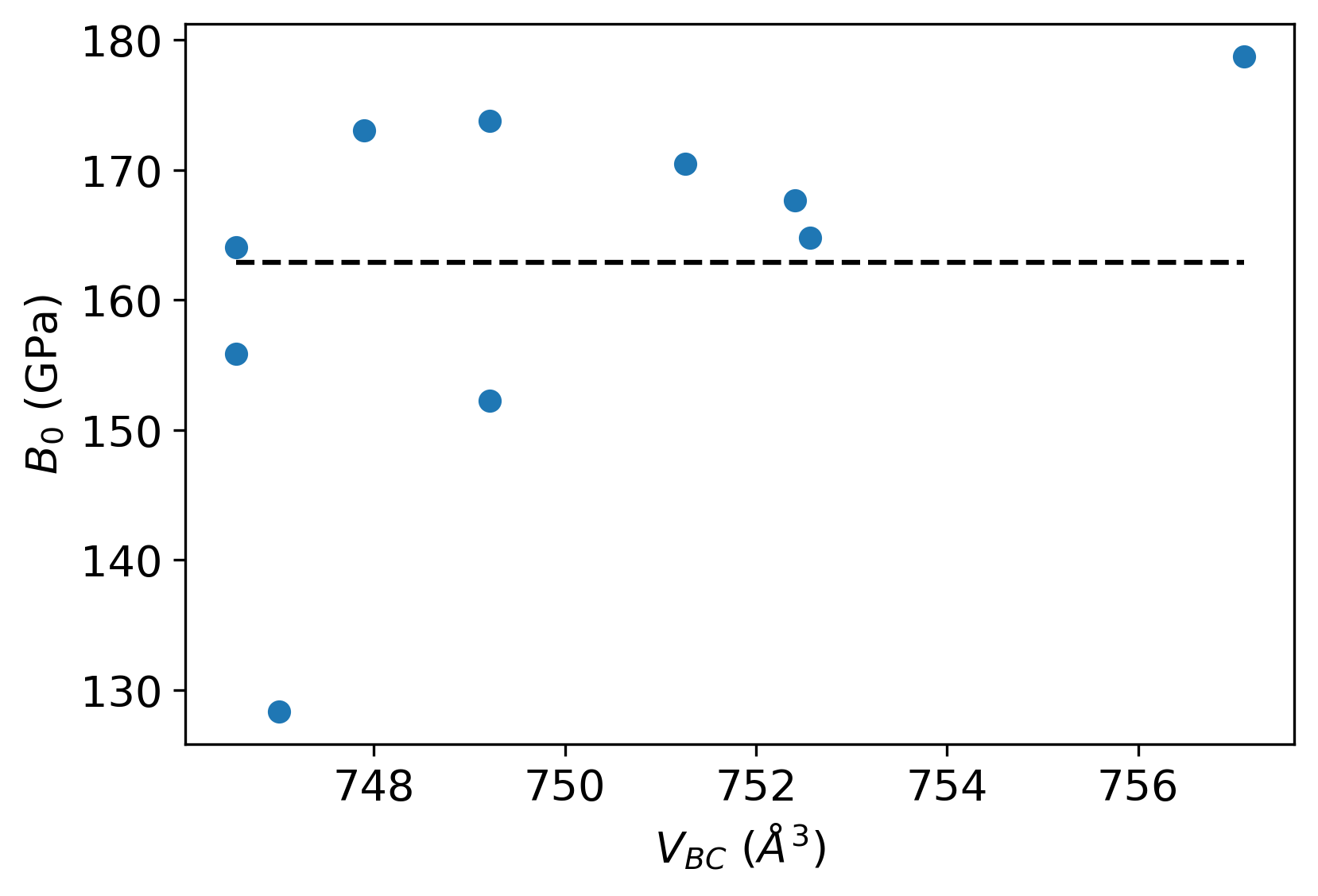}}
\end{tabular}
\caption{Zero-pressure bulk modulus calculated for all (a) crystalline and (b) amorphous supercells as a function of $V_{BC}$, representing stoichiometry, as defined in Eq.~\eqref{volume_correlation_formula}, where the lattice constants of elemental carbides are taken from \cite{yang_structural_2018}. The average bulk modulus is indicated by the dashed black line.}
\label{bulk_modulus_volume}
\end{figure*}

To summarise our results for all the crystalline and amorphous supercells energy dependence on volume, we present the average fit to the Birch-Murnaghan equation of state together with the standard deviations in Fig.~\ref{E_vs_V_all_supercells_combined}. The zero-pressure bulk modulus in the crystalline case is $B_{0}=(271.13 \pm 2.04)$ GPa. The standard deviation of our results is a factor hundred smaller than the average, therefore the influence of the differences in the stoichiometry is, as for the lattice constant, very limited in the crystalline case. Our result is comparable to other theoretical work: Y. Yang et al. \cite{yang_structural_2018} fitted with Eq. \eqref{BM}, resulting in $B_{0}=264.26$ GPa, whereas Q. Zhang \cite{zhang_understanding_2019} and P. Sarker et al. \cite{sarker_high-entropy_2018} calculated elastic constants and used these and the Voigt-Reuss-Hill approximation to show that $B=254$ GPa and $B=262$ GPa, respectively. Experimental work, in which (TaNbHfTiZr)C with roughly 10\% carbon vacancies is studied, found a bulk modulus of $(205 \pm 7)$ GPa using an ultrasonic pulser/receiver \cite{wen_thermophysical_2020}. The mismatch with our result can be attributed to the sample's porosity, in which case  nanohardness and elastic modulus measured with nanoindentation would have better  resembled the theoretical values than the  quantities measured with the ultrasonic pulser \cite{wen_thermophysical_2020}.

In the amorphous case, we calculated the zero-pressure bulk modulus to be $B_{0}=(162.89 \pm 13.83)$ GPa, with a standard deviation one order of magnitude larger than in the crystalline case. This can be understood by the large structural variations between the supercells for the amorphous configurations. 

Interestingly, the average bulk modulus in the amorphous case is much lower than in the crystalline case. This can be related to the fact that the equilibrium volume of the amorphous supercells is found to be larger with respect to the crystalline supercells, which would result in lower interaction between the atoms. Fig.~\ref{E_vs_V_all_supercells_combined} confirms the lower compressibility of the crystalline phase: a change in volume $V$ (e.g. from $V_{0}$ to  $V_{0} + 20 \, \text{\AA}\textsuperscript{3}$) leads to a larger increase in energy of the crystalline supercells compared to the corresponding amorphous ones.

Importantly, Fig. \ref{bulk_modulus_volume} shows that the bulk modulus and $V_{BC}$, quantifying the differences in stoichiometry, are not strongly correlated not only in the amorphous, but also in the crystalline case. This is in contrast to the lattice constant, where a stoichiometry dependence, although weak, was observed. Still, crystalline cells with the same stoichiometry have a similar bulk modulus.

\subsection{Electronic properties}

\begin{figure*}
\centering
\begin{tabular}{cc}
Crystalline & Amorphous \\
\includegraphics[width=0.45\textwidth]{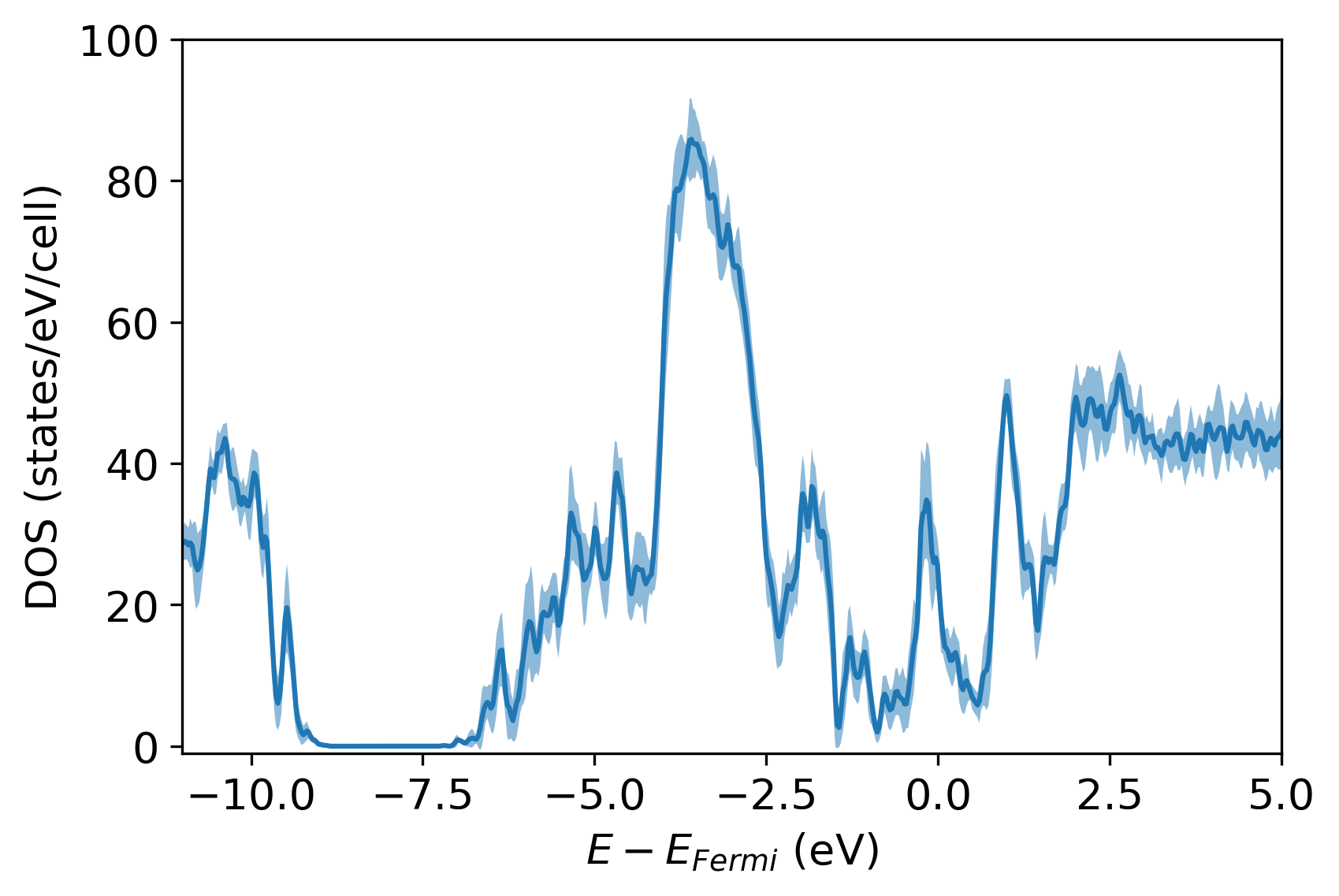} & \includegraphics[width=0.45\textwidth]{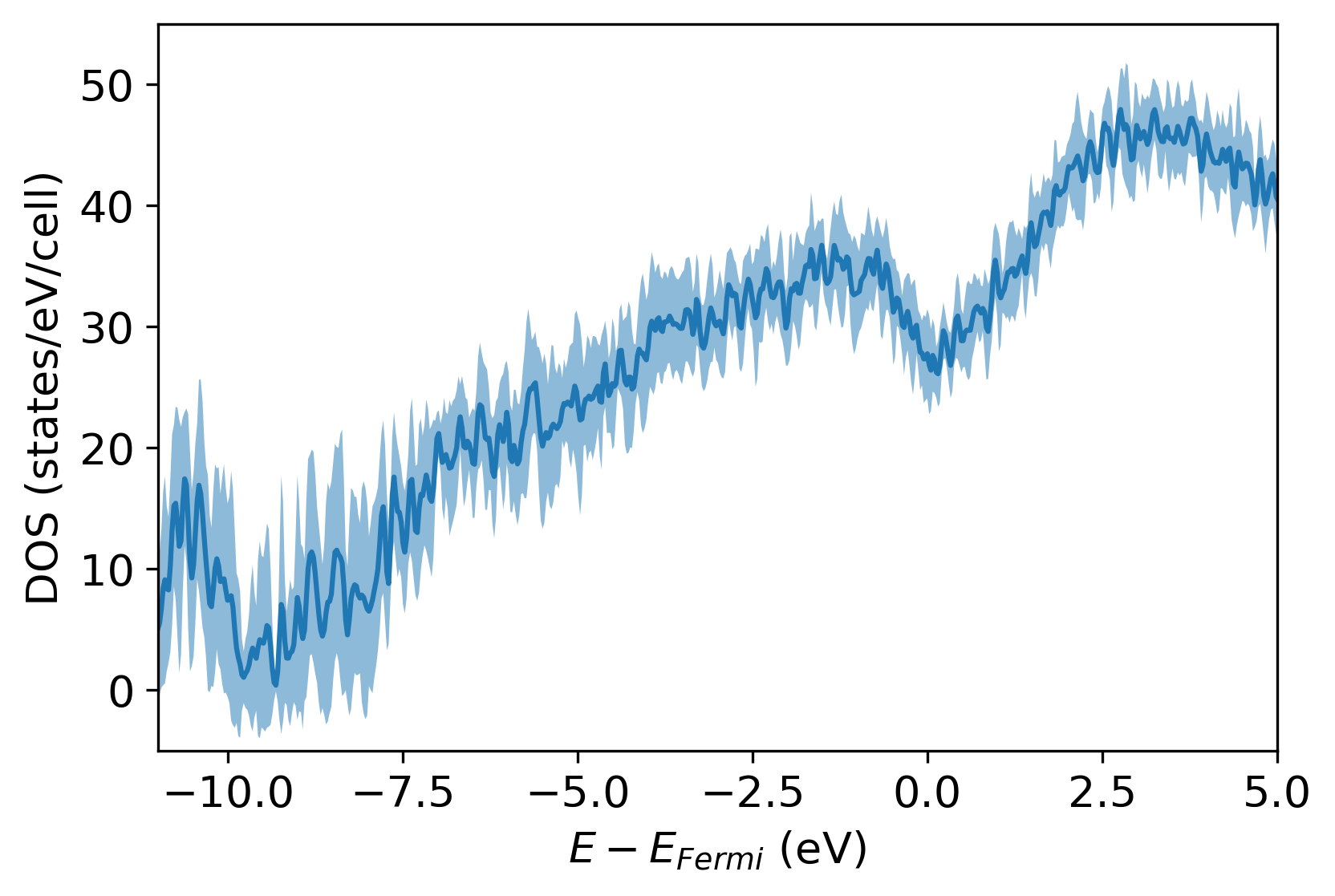} \\
\includegraphics[width=0.45\textwidth]{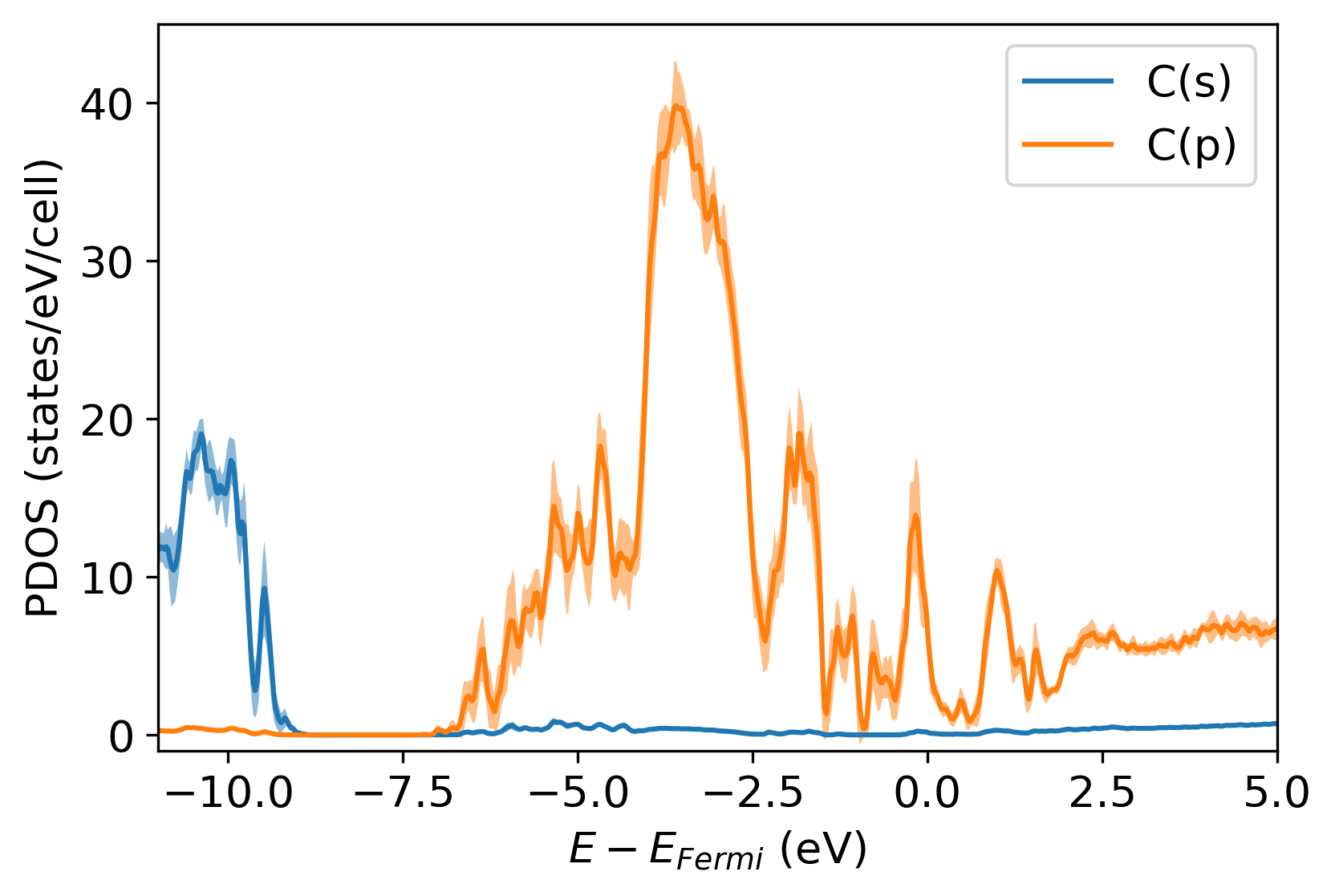} & 
\includegraphics[width=0.45\textwidth]{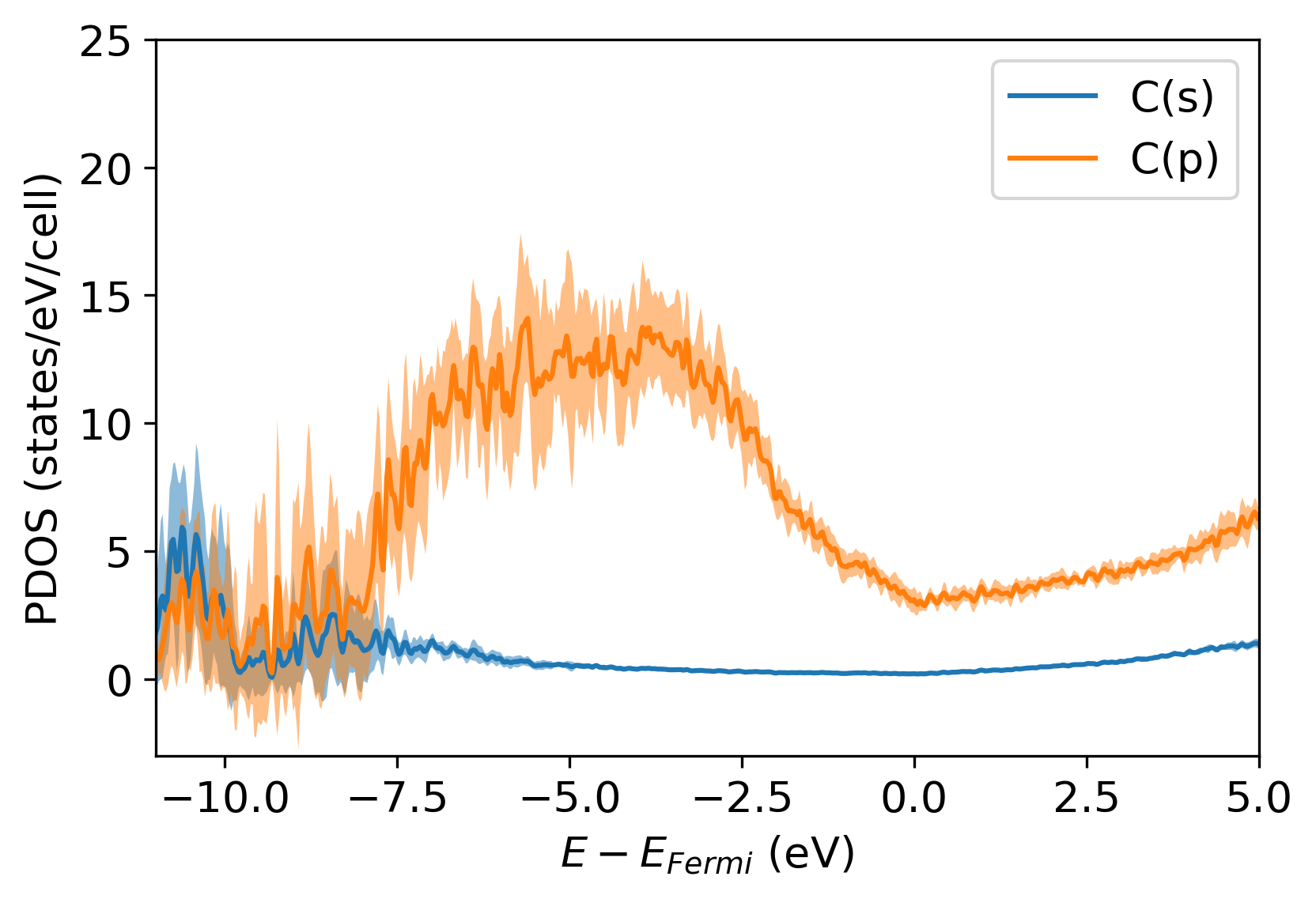}
\end{tabular}
\caption{DOS and projected DOS (PDOS) on the $s$ and $p$ orbitals of carbon for the crystalline and amorphous cells close to the Fermi energy $E_{Fermi}$. The solid line is the average of the (P)DOS over all the supercells. The standard deviation is shown by the wider coloured area.}
\label{PDOS_paper_part1}
\end{figure*}

\begin{figure*}
\centering
\begin{tabular}{cc}
Crystalline & Amorphous \\
\includegraphics[width=0.45\textwidth]{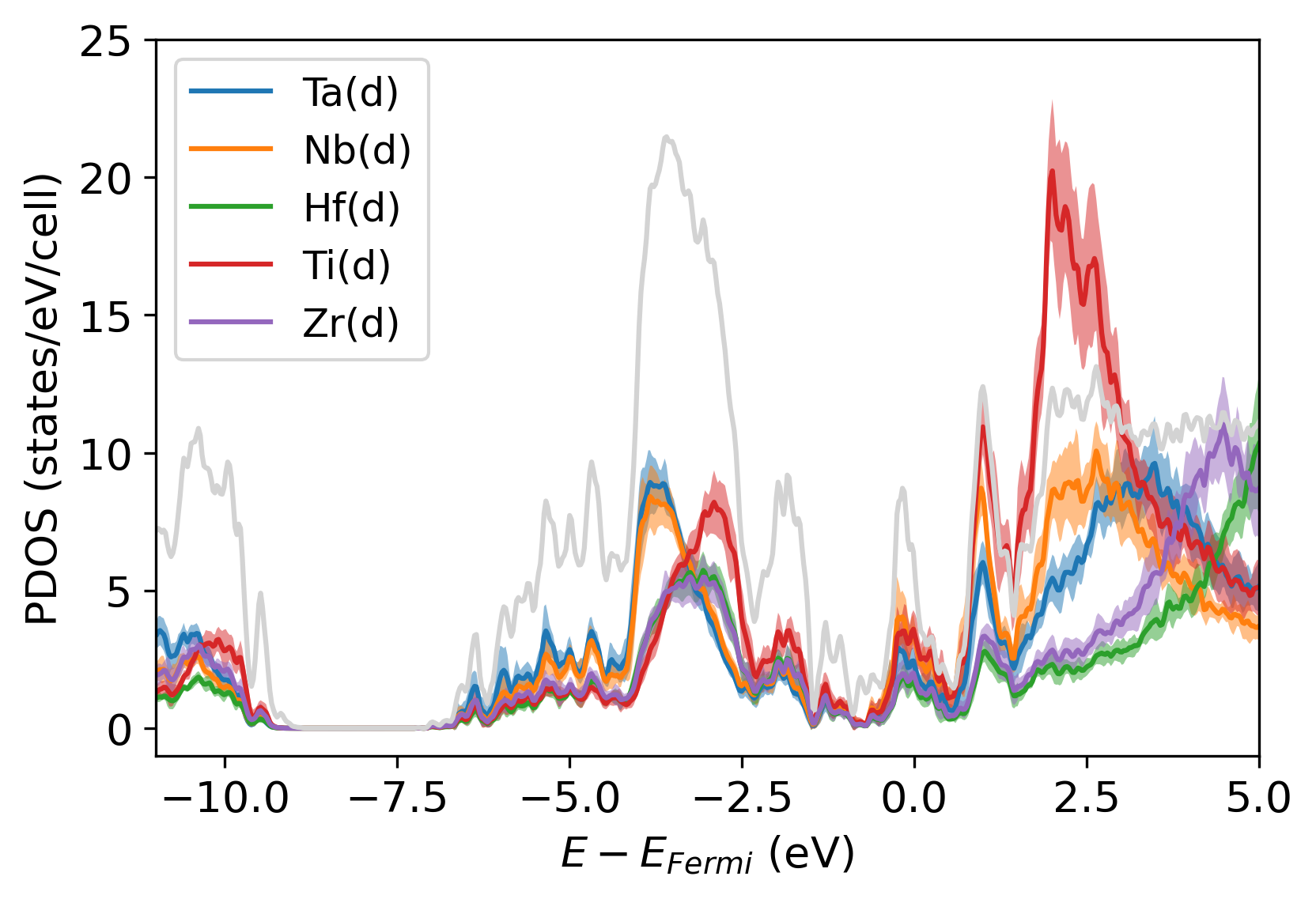} &
\includegraphics[width=0.45\textwidth]{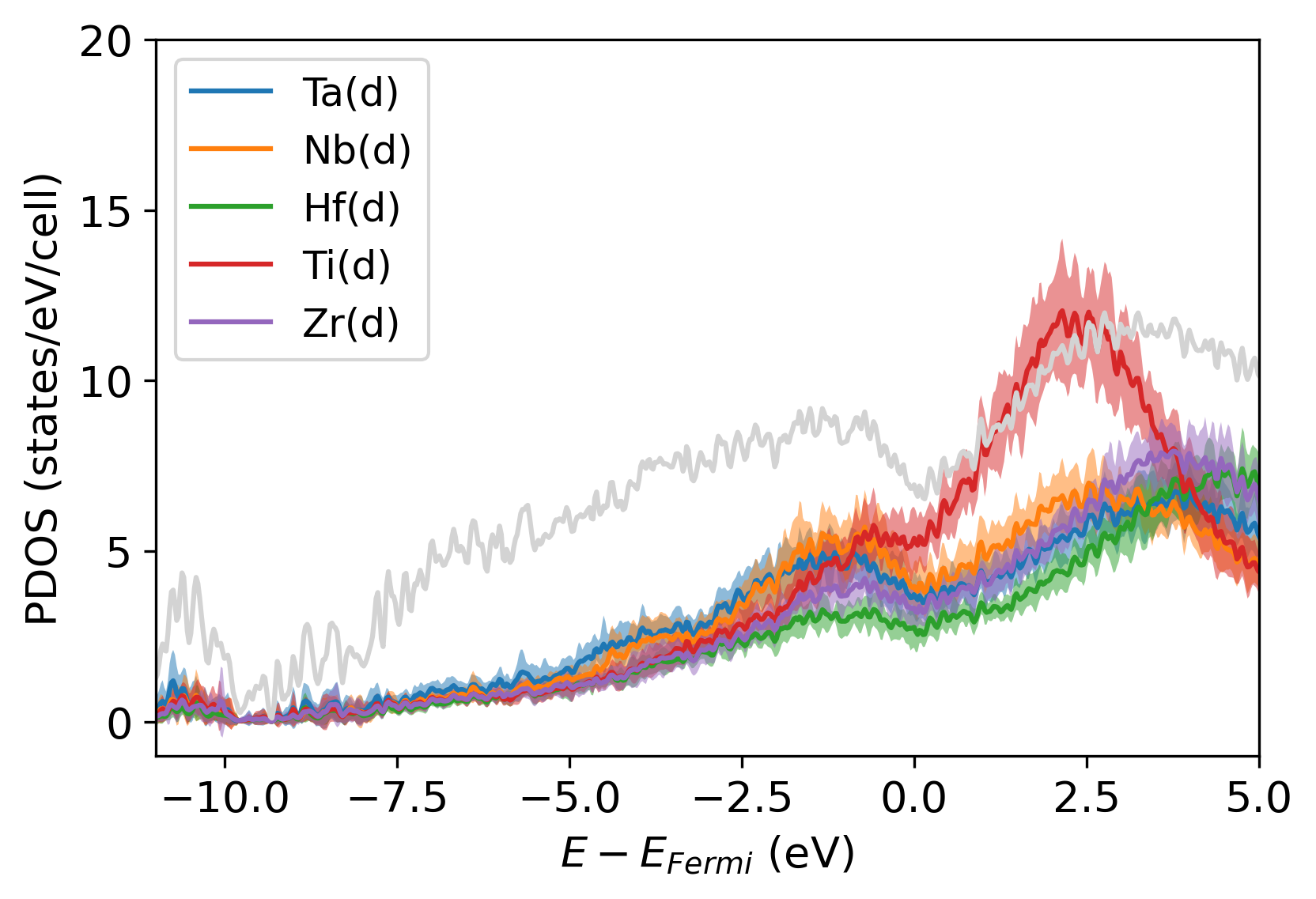} \\
\includegraphics[width=0.45\textwidth]{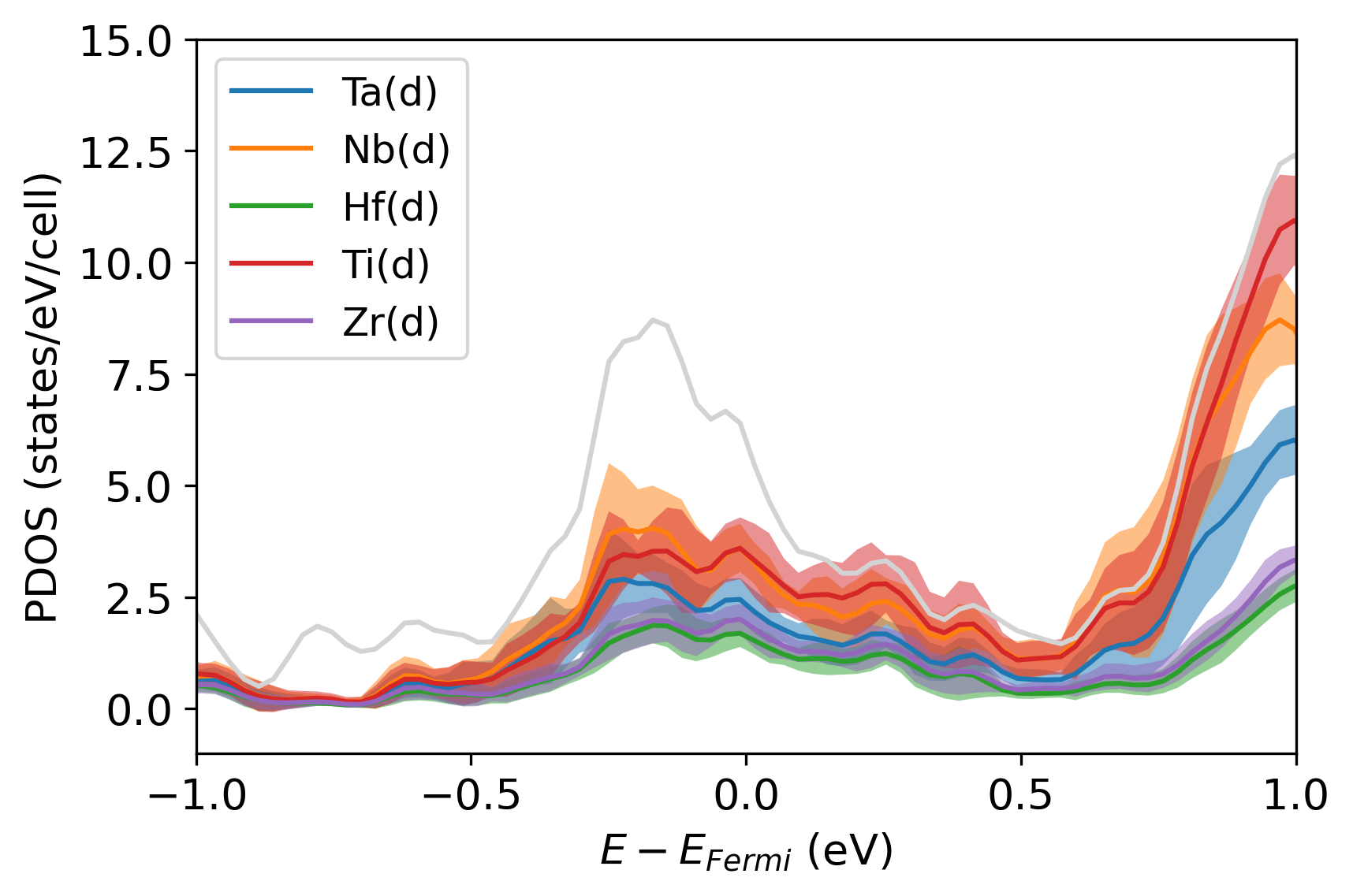} &
\includegraphics[width=0.45\textwidth]{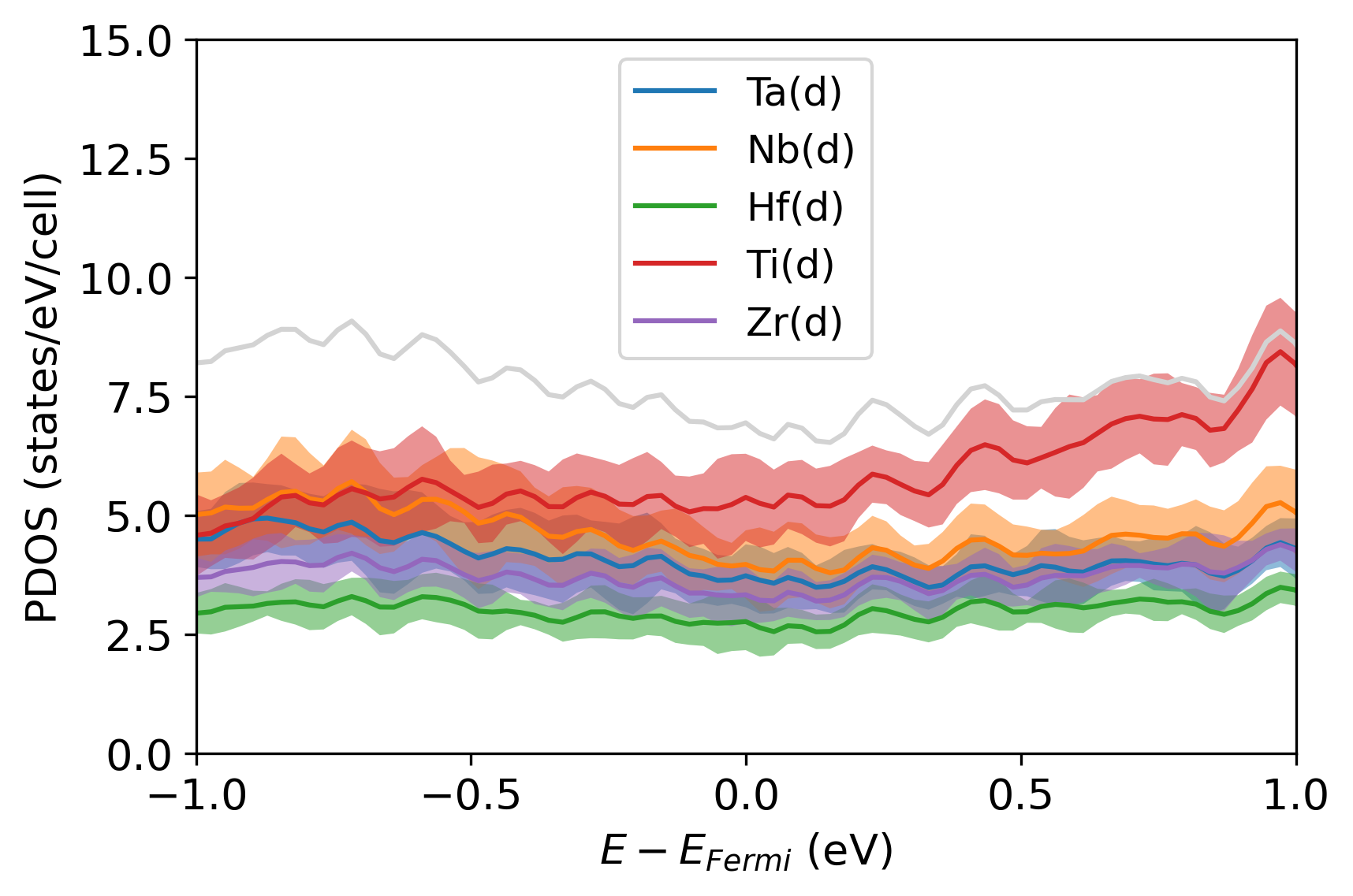} 
\end{tabular}
\caption{PDOS for the transition metal $d$-orbitals for the crystalline and amorphous cells close to the Fermi energy $E_{Fermi}$. The solid line is the PDOS averaged over all the respective supercells. The standard deviation is shown by the wider  coloured area. For comparison we plot the total DOS in gray (divided by 4).} 
\label{PDOS_paper_part2}
\end{figure*}

The DOS was calculated using DFT for all the crystalline and amorphous supercells. We also calculated its elemental and orbital contributions by calculating the projected DOS (PDOS). In particular, we evaluate the PDOS on the $s$ and $p$ orbitals of carbon atoms, and the $d$ orbitals of the transition metal atoms. Data for all the stoichiometries were averaged and are plotted together with their standard deviations shown in Figs.~\ref{PDOS_paper_part1} and \ref{PDOS_paper_part2}, separately for the crystalline and amorphous supercells.

We find that both the crystalline and amorphous phases are metallic. The DOS of the amorphous cells shows a dip close to the Fermi energy, but otherwise looks relatively featureless. The crystalline cells have a gap in the DOS from roughly -7 eV to -9 eV. This gap and the finite DOS at the Fermi energy in the crystalline case confirm theoretical calculations done in \cite{yang_structural_2018} and \cite{zhang_understanding_2019}.

From -7 eV to +5 eV, the contribution to the DOS of the crystalline phase is negligible for C($s$) whereas it is significant for C($p$). From -9 eV to -11 eV, the contributions are swapped. For the amorphous cells, the PDOS of C($p$) is much larger than C($s$) between -7.5 eV and +5 eV whereas both are indistinguishable from -7.5 eV to -11 eV.

For the $d$ orbitals of the transition metals, the PDOS is similar between all the elements, with a slight dominance of the $d$ orbital of titanium above the Fermi energy in both the crystalline and amorphous cases. This dominance is clearly more pronounced than the standard deviation and is therefore not a consequence of the details of the stoichiometry. This dominance of Ti($d$) and the significant contribution from C($s$) below -9 eV were also found in \cite{yang_structural_2018} and \cite{zhang_understanding_2019}, in the crystalline case.

\section{Discussion}

Previous work on disordered binary metal carbide systems has shown that their properties are tuneable. In the amorphous phase, the relative bond content determines hardness, resistivity and elasticity \cite{andersson_amorphousCrC_2012, furlan_amorphousFeC_2015, nedfors_Nb-B-Si_2014}. Amorphicity may lower cleavage fracture susceptibility and grain boundary decohesion \cite{kaloyeros_amorphous_1986}, which lowers corrosion through a reduced ability of oxygen to diffuse through grain boundaries. This can also be related to the flattening of the DOS as a function of energy that we calculated in the amorphous case and which can be attributed to the increased disorder in bond energy, as observed in a previous work on amorphous transition metal carbides \cite{kadas_structural_2012}.

In turn we show that amorphicity also lowers the bulk modulus as compared to the single-crystalline case. This implies that the structural disorder acts as a tuning knob between, on one side, the improvement of surface properties such as cleavage fracture susceptibility and corrosion, and hardness on the other side. This tuning knob can be adjusted depending on target applications. The bulk modulus was chosen in our comparative study because it is a measure of the elastic properties of a material that can easily be translated from ordered solids to liquids and disordered solids. The tunability of the bulk modulus with disorder is relevant to applications requiring high or low compressibility of the material or the storage of mechanical energy and suggests that also other mechanical properties such as hardness, Young’s modulus, or shear modulus depend on structural disorder in a comparable way.

The aforementioned dependence of material properties on bond distribution in the amorphous phase of metal carbides \cite{jansson_Nb-Si-C_2013, jansson_ZrSiC_2012} implies a range of tuneable material properties by varying the degree of disorder present in a metal carbide coating. Our work, calculating properties in the fully crystalline and fully amorphous limits, gives bounds on the available tuning range.

Apart from the change in flatness of the DOS, its value at the Fermi level is also strongly affected by the structural disorder. This can be related to experimental work showing a strong increase in resistivity with increasing relative amount of metal-to-carbon bonds in the amorphous phase \cite{jansson_ZrSiC_2012,jansson_Nb-Si-C_2013}. However, in these experiments, amorphicity is obtained by adding silicon to the system, which may be a further source of the raised resistivity, in addition to the inhibition of band formation by disorder. Nevertheless, it is not strictly necessary to introduce additional elements, such as Si \cite{jansson_ZrSiC_2012, jansson_Nb-Si-C_2013} or B \cite{nedfors_Nb-B-Si_2014}, to drive the amorphization in metal carbides, since  amorphized yet metalloid-free ternary metal carbides have been reported previously \cite{trindade_amorphousWC_1998, magnuson_amorphousCrC_2012, furlan_amorphousFeC_2015, olovsson_amorphousCrCXAS_2016}. In the specific case of transition metal carbide films, TEM-amorphous CrC \cite{magnuson_amorphousCrC_2012} and WC-based films \cite{trindade_amorphousWC_1998} have been synthesized and their electronic structure has been characterized experimentally. In this context, our calculations show that changes in electronic properties are at least partially intrinsic to variations in the level of structural disorder in carbides. The large change in the DOS at the Fermi level, for example, directly connects to experimental observations on transition metal carbides.

\section{Conclusion}

We calculated the lattice constant, bulk modulus and DOS of (TaNbHfTiZr)C in ten crystalline and ten amorphous configurations with slightly varying stoichiometries, using DFT and stochastic quenching. To model the random arrangement of transition metal atoms, we randomly generated $2\times2\times2$ supercells, with stoichiometries approaching the nominal one.

We found that the crystalline supercells have smaller relaxed lattice constants compared to amorphous supercells, and that the relaxed lattice constants for crystalline supercells are correlated with an average volume per transition metal specie, which can be used to quantify the influence of stoichiometry.

The bulk modulus depends strongly on the crystalline or amorphous nature of the crystal structure: it is $(271.13 \pm 2.04)$ GPa in the crystalline case and $(162.89 \pm 13.83)$ GPa in the amorphous case, making the crystalline phase of (TaNbHfTiZr)C much less compressible than its amorphous counterpart. This agrees with the larger relaxed volume of the amorphous supercells.

The DOS of the crystalline and amorphous phases of (TaNbHfTiZr)C is of metallic character, with all the species contributing to the DOS at the Fermi level. The DOS for crystalline configurations displays sharp features, including a small peak at the Fermi level, while the one for amorphous configurations is almost featureless.

All the structural, mechanical and electronic properties we calculated are only impacted in a limited way by variations between supercells, thus showing that $2\times2\times2$ supercells with approximate stoichiometries can be reliably used in \emph{ab-initio} calculations of materials properties of transition metal high-entropy carbides.

\section*{Acknowledgments}

We thank Edan Lerner for kindly providing the Lennard-Jones relaxation code. This work was carried out on the Dutch national e-infrastructure with the support of SURF Cooperative.

\section*{Appendix}

\begin{table}
\centering
\begin{tabular}{|c|c|}
\hline
Supercell & Ti:Zr:Nb:Hf:Ta \\ \hline
1 & 6 : 7 : 6 : 6 : 7                         \\\hline
2 & 7 : 6 : 7 : 6 : 6                     \\\hline
3 & 5 : 7 : 6 : 7 : 7                     \\\hline
4 & 7 : 6 : 6 : 6 : 7                         \\\hline
5 & 7 : 7 : 5 : 7 : 6                          \\\hline
6 & 7 : 6 : 6 : 6 : 7                             \\\hline
7 & 6 : 5 : 7 : 7 : 7                          \\\hline
8 & 7 : 6 : 6 : 7 : 6                      \\\hline
9 & 7 : 6 : 6 : 7 : 6                     \\\hline
Ref. cell \cite{yang_structural_2018} & 6 : 6 : 6 : 7 : 7                 \\\hline
\end{tabular}
\caption{The stoichiometries of the nine cells and the reference cell from \cite{yang_structural_2018}.}
\label{stoichiometries}
\end{table}

\begin{table*}[h]
\centering
\noindent\makebox[\textwidth]{ 
\begin{tabular}{|c|cccc|cccc|}
\hline
& \multicolumn{4}{|c|}{Crystalline} &  \multicolumn{4}{|c|}{Amorphous}\\\hline
Supercell & $B_{0}$ (GPa) & $B_{0}'$ (unitless) & $a_{0}$ (\AA) & $E_{0}$ (eV) & $B_{0}$ (GPa) & $B_{0}'$ (unitless) & $a_{0}$ (\AA) & $E_{0}$ (eV) \\\hline
1 & 270.98 & 4.15 & 4.52 & -51943.92 & 167.64 & 6.96 & 4.64 & -51890.32 \\
2 & 271.36 & 4.13 & 4.51 & -52213.99 & 128.34 & 1.73 & 4.68 & -52162.61 \\
3 & 271.92 & 4.12 & 4.53 & -51700.81 & 178.70 & 5.43 & 4.65 & -51650.33 \\
4 & 272.30 & 4.15 & 4.51 & -52245.23 & 164.07 & 3.82 & 4.65 & -52194.77 \\
5 & 266.93 & 4.11 & 4.52 & -51729.50 & 164.76 & 5.60 & 4.63 & -51680.77 \\
6 & 272.20 & 4.16 & 4.51 & -52245.16 & 155.86 & 7.95 & 4.67 & -52194.06 \\
7 & 275.11 & 4.14 & 4.51 & -52273.11 & 173.03 & 4.29  & 4.64 & -52222.79 \\
8 & 269.45 & 4.12 & 4.52 & -52000.80 & 152.22 & 6.96 & 4.68 & -51947.23 \\
9 & 269.69 & 4.13 & 4.52 & -52001.11  & 173.77 & 4.42 & 4.63 & -51949.08 \\
Y. Yang et al. \cite{yang_structural_2018} & 271.34 & 4.24 & 4.52 & -52001.80 & 170.45 & 5.23 & 4.65 & -51950.19 \\\hline
avg. & 271.13 & 4.14 & 4.519 & -52035.54 & 162.89  &  5.24 & 4.65 & -51984.21 \\
std.dev. & 2.04 & 0.03 & 0.006 & 198.45 & 13.83 & 1.71 & 0.02 & 198.31 \\\hline
\end{tabular}
}
\caption{Numerical results obtained by fitting with the third-order Birch-Murnaghan equation of state are given for all supercells, together with the averages (avg.) and standard deviations (std.dev.). All numbers are rounded down to two decimals, except for average and standard deviation of the crystalline lattice constant $a_{0}$: they are rounded such that the standard deviation has at least one significant number.}
\label{BM_EOS_all_fitted_values}
\end{table*}

\newpage
\bibliography{bibliography}

\end{document}